\documentclass[prl,twocolumn,amsmath,amssymb,graphicx]{revtex4-2}

\usepackage{bm,url,natbib,textcase,color,hyperref}

\newcommand{\arXiv}[1]{\href{http://www.arXiv.org/abs/#1}{arXiv:#1}}
\newcommand{\beq}{\begin{equation}}
\newcommand{\eeq}{\end{equation}}
\newcommand{\del}{\partial}

\def\s{{\textstyle \star}}

\begin{document}

{\phantom{.}\vspace{-2.5cm}\\\flushright Imperial-TP-KM-2021-02\\
}
\bigskip

\title{Democratic Lagrangians for Nonlinear Electrodynamics\vspace{-1mm}}

\author{Zhirayr Avetisyan}
\email{z.avetisyan@math.ucsb.edu}
\affiliation{Department of Mathematics, University of California, Santa Barbara CA 93106-3080, USA}
\affiliation{Regional Mathematical Center of Southern Federal University, Rostov-on-Don 344090, Russia\vspace{-3.5mm}}
\author{Oleg Evnin}
\email{oleg.evnin@gmail.com}
\affiliation{Department of Physics, Faculty of Science, Chulalongkorn University, Bangkok 10330, Thailand}
\affiliation{Theoretische Natuurkunde, Vrije Universiteit Brussel and International Solvay Institutes, Brussels 1050, Belgium\vspace{-3.5mm}}
\author{Karapet Mkrtchyan} 
\email{k.mkrtchyan@imperial.ac.uk}
\affiliation{Theoretical Physics Group, Blackett Laboratory, Imperial College London, SW7 2AZ, UK}

\begin{abstract}

We construct a Lagrangian for general nonlinear electrodynamics that features electric and magnetic potentials on equal footing. In the language of this Lagrangian, discrete and continuous electric-magnetic duality symmetries can be straightforwardly imposed, leading to a simple formulation for theories with the $SO(2)$ duality invariance. When specialized to the conformally invariant case, our construction  provides a manifestly duality-symmetric formulation of the recently discovered ModMax theory.
We briefly comment on a natural generalization of this approach to $p$-forms in $2p+2$ dimensions.

\end{abstract}

\maketitle

Nonlinear electromagnetic theories with actions of the form
\beq
S=\int {\cal L}(s,p) \,d^4x,\quad s\equiv\frac12F_{\mu\nu}F^{\mu\nu},\quad p\equiv\frac12F_{\mu\nu}\,
\s  F^{\mu\nu}
\label{nlingen}
\eeq
have surfaced historically in relation to topics as diverse as mitigating classical field divergences due to point charges \cite{BI,Plebanski:1970zz,Ts_reg,PTs}, induced interactions of photons via coupling to matter \cite{EH,BBBB,Adler,DH,Dunne}, effective low-energy description of open string theory \cite{FTs,Ts_rev} and construction of regular black holes \cite{BH1,BH2,BH3,BH4}. Here, $F_{\mu\nu}\equiv\del_\mu A_\nu - \del_\nu A_\mu$ and the Hodge star is defined by $\s F_{\mu\nu}=\epsilon_{\mu\nu\sigma\rho} F^{\sigma\rho}/2$, while most generally we view $\cal L$ as an arbitrary function.

There are many reasons one may want to recast (\ref{nlingen}) in a language where both the electric potential $A_\mu$ and its dual magnetic potential $B_\mu$ (satisfying $\s F_{\mu\nu}\equiv\del_\mu B_\nu - \del_\nu B_\mu$ in the free Maxwell theory) appear explicitly. For one thing, one may be interested in coupling this theory to both electrically and magnetically charged matter, as done for free theories in \cite{Zw,BS,LM,HTr}. Furthermore in many cases, such as the free Maxwell theory, or the celebrated Born-Infeld (BI) theory \cite{BI,FTs,Ts_rev,ZH}, the equations of motion are invariant under continuous rotations of $F$ into $\s F$ \cite{GZ,BB83,GZ2,GZ3,GR,Tse3,PSch,HKS,KT,KT1,IZ,BLM1,BLST1,Kosyakov,BLST2,Kuzenko,Bunster:2012hm}, and it is desirable to have an action principle that manifests this symmetry \cite{Zw,SchS,PST1,PST2,PST3,Rocek:1997hi,Berman,Nurmagambetov,BC,PST4,INZ,Henneaux:1988gg,Bunster:2011qp,Bunster:2010kwr}.

The purpose of this Letter is to give a democratic formulation in terms of electric and magnetic potentials to a general theory of the form (\ref{nlingen}) and to explore the properties of this formulation for special cases of current interest. In our construction, we shall rely on the approach of \cite{K,SOK} that, for free fields, provides a polynomial reformulation of the Pasti-Sorokin-Tonin (PST) theory \cite{PST1,PST2,PST3}. It turns out that this approach is  well-suited for including interactions, as we shall now demonstrate.

\vspace{1mm}
\noindent{\bf General formulation:} We start by recalling the formulation of \cite{K,SOK} for free fields:
\begin{equation}
{\cal L}_{Maxwell}=-\frac14 H^b_{\mu\nu}H^b{}^{\mu\nu}+\frac{a(x)}4\,\,\epsilon_{bc}\,\varepsilon^{\mu\nu\lambda\rho}\,F^{b}_{\mu\nu}\,Q^{c}_{\lambda\rho}\,,  \label{Maxwell}
\end{equation}
where $H^b_{\mu\nu}\equiv F^b_{\mu\nu}+a\,Q^b_{\mu\nu}$, $b=1,2$, $a(x)$ is an auxiliary scalar familiar from the PST theory, and
\begin{equation}
F^b_{\mu\nu}=\partial_{\mu}\,A^b_{\nu}-\partial_{\nu}\,A^b_{\mu}\,,\quad Q^b_{\mu\nu}=\partial_{\mu}\,R^b_{\nu}-\partial_{\nu}\,R^b_{\mu}\,.
\end{equation}
Despite the large number of fields ($A^b$, $R^b$, $a$), this Lagrangian describes a single propagating Maxwell field whose electric potential is $A^1$ and magnetic potential $A^2$. To see this, we observe that (\ref{Maxwell}) is invariant under the usual gradient shifts of $A^b$ and $R^b$ (the Lagrangian depends only on the corresponding field strengths) and has two additional symmetries:
\beq
\delta a=0,\quad \delta A^b_\mu= -a u^b \,\del_\mu a,\quad \delta R^b_\mu=u^b\, \del_\mu a, 
\label{dashift}
\eeq
where $u^b(x)$ is an arbitrary doublet of scalar field parameters, and another gauge symmetry that shifts $a(x)$ arbitrarily and correspondingly corrects the other fields. Further details can be found in \cite{SOK}. As a consequence of these symmetries, any solution of the equations of motion can be gauged to
\begin{eqnarray}
    R^b=0,\qquad \s F^a_{\mu\nu}+\epsilon^{ab}\, F^b_{\mu\nu}=0\,,\label{DSM}
\end{eqnarray}
leaving a single propagating Maxwell field.

To include interactions in this formalism, we intend to deform the free action \eqref{Maxwell} in a way that maintains its gauge symmetries. To this end, we first note that $H^b_{\mu\nu}$ is invariant by itself under the transformations (\ref{dashift}) while the second term in (\ref{Maxwell}) changes by a total derivative. As a result, if we replace $H^b_{\mu\nu}H^b{}^{\mu\nu}$ by an arbitrary scalar function of  $H^b_{\mu\nu}$, the resulting interacting theory still automatically respects both gradient shifts of $A^b$ and $R^b$, and (\ref{dashift}), which is crucial for the emergence of (\ref{DSM}). 

There are six functionally independent scalars one can build from $H^b_{\mu\nu}$ ($U^{ab}=U^{ba}$, $V^{ab}=V^{ba}$):
\begin{equation}
    U^{ab}\equiv\frac12\,H_{\mu\nu}^a\,H^b{}^{\mu\nu}\,,\quad V^{ab}\equiv\frac12\,H_{\mu\nu}^a\,\s{H}^{b}{}^{\mu\nu}\,.\label{UV}
\end{equation}
Our algorithm to construct a nonlinear generalization of (\ref{Maxwell}) is then to start with
\begin{equation}
    {\cal L}= a\,\epsilon_{bc}F^b\wedge Q^c+f(U,^{\hspace{-0.7mm}11}U,^{\hspace{-0.7mm}12}U,^{\hspace{-0.7mm}22}V,^{\hspace{-0.5mm}11}V,^{\hspace{-0.5mm}12}V^{22})\,,\label{NLE}
\end{equation}
and constrain $f$ by the requirement that there is an extra symmetry shifting  $a(x)$ arbitrarily, so that it is a pure gauge degree of freedom, or equivalently \cite{SOK},
that the equations of motion for $A^b$ and $R^b$,
\begin{eqnarray}
&d[(f^U_{bc}+f^U_{cb})\,\s H^c-(f^V_{bc}+f^V_{cb})\,H^c+a\,\epsilon_{bc}\,Q^c]=0\,,\hspace{3mm}\label{Eq:A}\\
&d[a\{(f^U_{bc}+f^U_{cb})\,\s H^c-(f^V_{bc}+f^V_{cb})\,H^c-\epsilon_{bc}\,F^c\}]=0\,,\hspace{3mm}\label{Eq:R}
\end{eqnarray}
where $f^U_{ab}\equiv\del f/\del U_{ab}$, $f^V_{ab}\equiv\del f/\del V_{ab}$ ($f^U_{21}\equiv 0\equiv f^V_{21}$),
imply the equation of motion for $a(x)$,
\beq
Q^b\wedge\,K_b=0\,,\label{Eq:a}
\eeq
with
\beq
    K_a\equiv(f^U_{ab}+f^U_{ba})\,\s H^b-(f^V_{ab}+f^V_{ba})\,H^b-\epsilon_{ab}\,H^b\,.\label{conjecture4d}
\eeq
For that, we first note that multiplying (\ref{Eq:A}) by $a(x)$ and subtracting it from (\ref{Eq:R}), one gets (we use the differential form notation following the derivations of \cite{SOK})
\beq
    da\wedge K_b=0\,.\label{Kda}
\eeq
There is a natural way to ensure that (\ref{Kda}) implies (\ref{Eq:a}) in a manner analogous to the free theory \cite{SOK}. Indeed, if
\beq
    K_a\pm \epsilon_{ab}\,\s K_b\equiv 0\,,\label{Ksd}
\eeq
then (\ref{Kda}) implies $K_b=0$ by elementary differential form algebra. Hence, (\ref{Eq:a}) is satisfied whenever (\ref{Eq:A}-\ref{Eq:R}) are satisfied \footnote{We do not have a proof that \eqref{Ksd} is the only way to ensure the gauge symmetry that shifts $a$, but it is satisfactory in that it allows us to construct a sufficiently large class of theories.}.

One can translate (\ref{Ksd}) to the following condition on $f$:
\beq
    \pm \,\delta^{ac}\,(f^U_{cb}+f^U_{bc})-\epsilon^{ac}\,(f^V_{cb}+f^V_{bc})+\delta^a_b=0\,.\label{MatrixEq}
\eeq
These linear PDEs are solved in full generality by
\begin{eqnarray}
    &f(U,V)=\mp\tfrac12 U_{aa} + g(\lambda_1,\lambda_2)\,,\label{solutionNLE}\\
    &\lambda_1=\pm  U_{12}-\tfrac12\,(V_{11}-V_{22})\,,\label{lambda1}\\
    &\lambda_2=V_{12}\pm \tfrac12\,(U_{11}-U_{22})\,.\label{lambda2}
\end{eqnarray}
where $g(\lambda_1,\lambda_2)$ is an arbitrary function (we will henceforth use the upper signs only). The Lagrangian is then
\beq
    {\cal L}={\cal L}_{Maxwell}+g(\lambda_1,\lambda_2)\,,\label{NLED}
\eeq
where $\lambda_{1,2}$ can be read off (\ref{lambda1}-\ref{lambda2}) and (\ref{UV}).
This Lagrangian respects a gauge symmetry that shifts $a$:
\beq
    \delta a=\varphi(x)\,,\; \delta A^b_{\mu}=\!-a\,\delta R^b_{\mu}=\frac{\varphi}{(\partial a)^2}a\partial^\nu\! a(Q^b_{\nu\mu}-\epsilon^{bc}\!\star Q^c_{\nu\mu})\,.
\label{symashift}
\eeq
The gauge transformation rules do not depend on the function $g(\lambda_1,\lambda_2)$ and, in particular, are the same as in the free case \cite{SOK}
; $\lambda_{1,2}$ are invariant under \eqref{dashift} and \eqref{symashift}.
The Lagrangian (\ref{NLED}) contains a single arbitrary function of two variables, as does (\ref{nlingen}).
We shall proceed to show its relation to the single-field formulation (\ref{nlingen}) after discussing the surprisingly simple way additional electric-magnetic symmetries can be imposed on (\ref{NLED}).

\vspace{1mm}
\noindent{\bf Duality symmetry:} Under the discrete $Z_4$ interchange of electric and magnetic degrees of freedom, $H^1\to H^2\,, H^2\to -H^1$, and hence
$\lambda_1\to -\lambda_1, \lambda_2\to -\lambda_2$.
Therefore, theories with such discreet duality symmetry are described by $g(\lambda_1,\lambda_2)$ satisfying
\beq
    g(-\lambda_1,-\lambda_2)=g(\lambda_1,\lambda_2)\,.\label{DiscreteDuality}
\eeq
Next, one may ask for the full $SO(2)$ duality symmetry with respect to rotating $A^b$ and $R^b$ in the $b$-plane. Under such a rotation by an angle $\alpha$, the pair $(\lambda_1,\lambda_2)$ simply rotates as a vector by an angle $2\alpha$, therefore only the radial part is invariant. Hence, the $SO(2)$-invariant theories are encoded in full generality by
\beq
    g(\lambda_1,\lambda_2)=h(w)\,,\qquad w=\sqrt{\lambda_1^2+\lambda_2^2}\,,\label{hw}
\eeq
and their Lagrangian is given by 
\beq
    {\cal L}={\cal L}_{Maxwell}+h(w)\,.\label{GDST}
\eeq
One can show that
\beq
    w=\sqrt{ - \det{\cal H}}\,,\label{w}
\eeq
where ${\cal H}^{ab}\equiv  (\s H^a_{\mu\nu}-\epsilon^{ac} H^c_{\mu\nu})(\s H^b{}^{\mu\nu}-\epsilon^{bd} H^d{}^{\mu\nu})/2\,$.

\vspace{1mm}
\noindent{\bf Conformal invariance:} Another symmetry one may require from (\ref{nlingen}) is conformal invariance \cite{BLST1,Kosyakov,BLST2}, which can be expressed by a condition of the form
\beq
    U^{ab} f^U_{ab}+V^{ab}f^V_{ab}=f\,.
\eeq
Then, $g(\lambda_1,\lambda_2)$ 
is a homogeneous function of degree one:
\beq
    g=\lambda_1\, \tilde g(\lambda_1/\lambda_2)\,,
\eeq
where $\tilde g(x)$ is an arbitrary function.

If we require both $SO(2)$ symmetry and conformal invariance, (\ref{GDST}) reduces to
\beq
    {\cal L}=
    -\frac12\,H^b\wedge \star H^b+a\,\epsilon_{bc}F^b\wedge Q^c+\delta\, w\,,
\label{Lagconf}
\eeq
where $\delta$ is an arbitrary real number, $w$ given by (\ref{w}).
As we shall show below, this construction provides an explicit duality-symmetric formulation of the ModMax theory recently introduced in \cite{BLST1} (see also \cite{Kosyakov,BLST2, Kuzenko}). 

\vspace{1mm}
\noindent{\bf Analysis of the equations of motion:} As explained under \eqref{Ksd}, equations (\ref{Eq:A}-\ref{Eq:R}) imply
\beq
    K_b=0\label{K}
\eeq
for theories of the form (\ref{NLED}).
Plugging \eqref{K} into \eqref{Eq:A}, we deduce:
\beq
    da\wedge dR^b=0\,,
\eeq
exactly the same equation as for the free field case \cite{SOK}. This, in turn, implies that the fields $R^b$ can be gauge-transformed to zero (see \cite{SOK} for details):
\beq
    R^b=0\,,\qquad H^b=F^b\,.\label{R0}
\eeq
Note that the two equations \eqref{K} are Hodge-dual to each other due to \eqref{Ksd}, and can hence be expressed as a single equation ($g_1\equiv{\del g}/{\del\lambda_1}, g_2\equiv{\del g}/{\del\lambda_2}$):
\beq
\s F^1+F^2=\,g_2\,(\s F^1-F^2)-\,g_1\,\s(\s F^1-F^2)\,,\label{FF}
\eeq
which includes the free case \eqref{DSM} given by $g(\lambda_1,\lambda_2)=0$.
Taking into account that (with $R^b=0$)
\beq
\lambda_1=\tfrac12\,G_{\mu\nu}\,\s G^{\mu\nu}\,,\quad
\lambda_2=-\tfrac12\,G_{\mu\nu}\,G^{\mu\nu}\,,
\eeq
where $G_{\mu\nu}\equiv\s F^1_{\mu\nu}-F^2_{\mu\nu}$, $g(\lambda_1,\lambda_2)$ is expressed through $G$, while the right hand side of \eqref{FF} is equal to $-{\partial g}/{\partial G}$. Therefore, \eqref{FF} is an expression for $\s F^1+F^2$ in terms of $\s F^1-F^2$. This means that,
similarly to the free field case, there is only one independent field strength, while all the auxiliary fields have been gauged away. Hence, we are really dealing with a theory of one dynamical gauge field, as intended. 

\vspace{1mm}
\noindent{\bf Relation to the conventional single-field  formulation:} 
The two formulations of nonlinear electrodynamics, given by \eqref{nlingen} and \eqref{NLED}, are related by nonlinear algebraic equations involving derivatives of the corresponding Lagrangians. We now proceed to establish this relation. If one solves (\ref{FF}) for $F_1$, by Lorentz invariance, the resulting expression must be of the form (see, e.g., \cite{BLM1})
\beq
    F^1=\alpha(s,p) F^2 + \beta(s,p) \s F^2\,,\label{F1F2}
\eeq
where $s$ and $p$ are the two independent invariants of $F^2$,
\beq
   s=\frac12 F^2_{\mu\nu}F^2{}^{\mu\nu}\,,\quad p=\frac12 F^2_{\mu\nu}\s F^2{}^{\mu\nu},
\eeq
and $\alpha$ and $\beta$ depend on the specific form of $g$. If we apply the exterior derivative operator to (\ref{F1F2}), the left-hand side vanishes, and $F^1$ disappears from the equation. What remains is the equation of motion for the theory (\ref{nlingen}), where $F$ is identified with $F^2$ and
\beq
    \alpha(s,p)=-\frac{\partial{\cal L}}{\partial p}\,,\qquad \beta(s,p)=\frac{\partial{\cal L}}{\partial s}\,.
    \label{lagdev}
\eeq
Thus, our democratic formulation reproduces dynamically the theory (\ref{nlingen}) for a single electromagnetic field with self-interactions.

One can recast (\ref{F1F2}) as the following equations in terms of the invariants:
\beq
    g_1=\frac{2\,\alpha}{\alpha^2+(\beta+1)^2}\,,\qquad g_2=\frac{\alpha^2+\beta^2-1}{\alpha^2+(\beta+1)^2}\,.\label{galphabeta}
\eeq
Here, $g$ is specified as a function of $\lambda_1$ and $\lambda_2$. The latter can be extracted from (\ref{F1F2}) and (\ref{lambda1}-\ref{lambda2}) as
\beq
\begin{split}
    \lambda_1=2\,\alpha\,(1+\beta)\,s-[\alpha^2-(1+\beta)^2]\,p\,,\\
\lambda_2=[\alpha^2-(1+\beta)^2]\,s+2\,\alpha\,(1+\beta)\,p\,,\label{Lambdas}
\end{split}
\eeq
so that
\beq
    w\equiv\sqrt{\lambda_1^2+\lambda_2^2}=(\alpha^2+(\beta+1)^2)\,\sqrt{s^2+p^2}\,.\label{walphabetasp}
\eeq
For any concrete $g(\lambda_1,\lambda_2)$, (\ref{galphabeta}) provides a $2\times 2$ system of nonlinear algebraic equations for $\alpha$ and $\beta$ as functions of $s$ and $p$. Given a solution of this system, one can reduce the equations of motion of the democratic theory to the single-field form (\ref{F1F2}). Conversely, to recast a given single-field theory of the form (\ref{nlingen}) in terms of the democratic formulation (\ref{NLED}), one needs to obtain $\alpha$ and $\beta$ from (\ref{lagdev}), then $\lambda_1$ and $\lambda_2$ from (\ref{Lambdas}) and then $g$ from (\ref{galphabeta}).

\vspace{1mm}
\noindent{\bf Single-field formulation for duality-symmetric theories:} There is little one can say in general about solutions to (\ref{galphabeta}-\ref{Lambdas}), but extra structures emerge for special classes of theories. For the $SO(2)$-invariant case of \eqref{GDST}, one gets
\begin{equation}
\frac{\lambda_1}{w}h'=\frac{2\,\alpha}{\alpha^2+(\beta+1)^2}\,,\quad \frac{\lambda_2}{w}h'=\frac{\alpha^2+\beta^2-1}{\alpha^2+(\beta+1)^2}\,.\label{gabso2}
\end{equation}
Eliminating $h'$ gives $\lambda_1(\alpha^2+\beta^2-1)=2\alpha\lambda_2$, and then
\begin{equation}
\beta^2+\frac{2s}p\alpha\beta-\alpha^2=1,
\label{SO2symm}
\end{equation}
which is, in view of (\ref{lagdev}), exactly the same as the general $SO(2)$-invariance condition in the single-field formalism \cite{GZ,GR,HKS,BC,IZ,Kosyakov,BLST2}.
The remaining equation can then be written as
\begin{equation}
    (\alpha s+(\beta+1)p)\,\,h'\Big|_{w=\sqrt{s^2+p^2}(\alpha^2+(\beta+1)^2)}=\alpha\sqrt{s^2+p^2}.
\label{eqSO2albe}
\end{equation}
If one is moving from the single-field formalism to the democratic one for $SO(2)$-invariant theories, (\ref{SO2symm}) is satisfied from the start, and (\ref{eqSO2albe}) is what must be solved to reconstruct $h(w)$.

\vspace{1mm}
\noindent{\bf The ModMax theory:}
For the conformally invariant case (\ref{Lagconf}), equation (\ref{eqSO2albe}) becomes linear, and it can be solved together with (\ref{SO2symm}):
\begin{align}
 & \alpha(s,p)=-\sinh{\gamma} \frac{p}{\sqrt{p^2 + s^2}}\,,\\
 &\beta(s,p)=\sinh{\gamma}\frac{s}{\sqrt{p^2 + s^2}} - \cosh{\gamma}\,.
\end{align}
With these functions, (\ref{F1F2}) reproduces the equations of motion of the ModMax theory introduced in \cite{BLST1} and defined by the Lagrangian
\beq
    L(s,p)=-\cosh{\gamma}\,s+\sinh{\gamma}\sqrt{s^2+p^2}\,.\label{ModMax}
\eeq
Thus, with the identification
\beq
    \delta=\coth{\frac{\gamma}{2}}\,,
\eeq
a democratic description of the ModMax theory is provided by (\ref{Lagconf}).
This solution corresponds to $\delta > 1$ ($\gamma>0$) or $\delta <-1$ ($\gamma<0$). For the region $-1<\delta<1$, which can be given as $\delta=\tanh{\tfrac{\gamma}{2}}$, we get the same action \eqref{ModMax}, but with an overall minus sign. This is what happens if we replace $F_{\mu\nu}\to \s F_{\mu\nu}\,,\; \gamma\to -\gamma$ in the Lagrangian \eqref{ModMax}, so it is the ModMax theory written in terms of the dual magnetic potential. 
The case $\delta=1$ 
\footnote{For $\delta=1$, one cannot resolve (\ref{FF}) into (\ref{F1F2}). Rather, (\ref{FF}) implies that
$\mathbf{V}=0$, $\det\mathbf{U}=0$, where $\mathbf{V}$ and $\mathbf{U}$ are the $2\times 2$ matrices defined by (\ref{UV}) with $H^b\to F^b$. These equations manifest enhanced $SL(2,R)$ symmetry. One can then always apply a duality rotation, so that $U_{11}=0=V_{ab}$. Thus, the electric field strength satisfies $s=0=p$ in the notation of (\ref{nlingen}), known as the equations of motion of the Bia\l ynicki-Birula (BB) electrodynamics \cite{BB,Chruscinski} (see also \cite{BLST2}). This theory does not admit a single-potential Lagrangian formulation, while it arises naturally from the democratic Lagrangian we consider here, highlighting the power of our formalism.} corresponds to the Bia\l ynicki-Birula electrodynamics \cite{BB,Chruscinski}.

\vspace{1mm}
\noindent{\bf Other duality-invariant theories:} 
While we have outlined the general way to connect the single-field and democratic formulations, it involves solving nonlinear algebraic equations, which is only possible explicitly in special cases. Furthermore, the functions appearing in the democratic Lagrangian may be complicated for known simple single-field theories, and vice versa.

We demonstrate here the conversion  procedure for the generalized BI theory \cite{BLST1,BLST2},
\beq
    L_{GBI}=\sqrt{UV}-T,\quad U\equiv 2u + e^\gamma\, T,\quad V\equiv - 2v + e^{-\gamma}\, T\,,\label{GBI}
\eeq
where $u\equiv(s + \sqrt{p^2 + s^2})/2$, $v\equiv(-s + \sqrt{p^2 + s^2})/2$, and $T$ is an arbitrary constant (related to the string tension in \cite{FTs}).
The BI theory corresponds to $\gamma=0$.
From \eqref{gabso2},
\beq
    h'=\frac{\sqrt{U}-\sqrt{V}}{\sqrt{U}+\sqrt{V}}\,.
\label{hprimeGBI}
\eeq
From \eqref{Lambdas}, 
\beq
    w=\frac{T}{2}\Big(e^{-\gamma}\frac{U}{V}-e^{\gamma}\Big)\Big(1+\sqrt{\frac{V}{U}}\Big)^2\,.
\eeq
Introducing $e^\lambda\equiv \sqrt{V/U}$, one can multiply (\ref{hprimeGBI}) by $\del w/\del\lambda$ and integrate to obtain
\beq
\begin{split}
 &h(\lambda)=
 4\, T\, \sinh^2\frac{\lambda}2\,\, \cosh(\lambda+\gamma)\,,\\
 &w(\lambda)=-4\, T\, \cosh^2\frac{\lambda}2\,\, \sinh(\lambda+\gamma)\,.
\end{split}
\eeq
These formulas provide an implicit definition for $h(w)$ corresponding to the generalized BI theory in the democratic formulation (\ref{GDST}).

\vspace{1mm}
\noindent{\bf Conclusions:}
We have provided a democratic formulation (\ref{NLED}) for the nonlinear electrodynamics (\ref{nlingen}) that explicitly features both electric and magnetic gauge potentials. It includes two auxiliary gauge fields and an auxiliary scalar, all of which are pure gauge degrees of freedom. The propagating degrees of freedom are those of (\ref{nlingen}).

The $SO(2)$ duality invariance is expressed in this formalism by the strikingly simple condition (\ref{hw}) which takes place of the nonlinear PDE given by (\ref{SO2symm}) and (\ref{lagdev}) responsible for the same property in the single-field formalism (this PDE is sometimes referred to \cite{IZ,BC} as the Courant-Hilbert equation after the classic treatise \cite{CH}). This makes it easy to specify arbitrary $SO(2)$-invariant interactions in (\ref{GDST}), without any need to satisfy additional constraints. In particular, polynomial interactions can be straightforwardly introduced  by choosing $h(w)$ in (\ref{GDST}) as a polynomial in $w^2$. 
If conformal symmetry is imposed in addition to the $SO(2)$ invariance, the theory further simplifies to (\ref{Lagconf}), which is a democratic formulation of the ModMax theory introduced in \cite{BLST1}.

Our formulation has been developed as a nonlinear extension of the approach employed for free fields in \cite{K,SOK}. In the free field context, this approach is closely related to the PST formulation \cite{PST1,PST2,PST3}, which is recovered by integrating out the auxiliary form fields. The relation is less obvious for the interacting theories discussed here since the equations of motion are no longer linear. If the auxiliary gauge fields can be successfully integrated out, a PST-like formulation of nonlinear electrodynamics will be produced. 

Our approach to constructing the Lagrangian (\ref{NLED}) has been rather systematic in that we started with a simple Lagrangian ansatz that automatically respects all the necessary gauge symmetries except for the symmetry that shifts the auxiliary scalar. Then, enforcing this last symmetry fixed the form of the Lagrangian. An advantage of the resulting theory (\ref{NLED}) is that all the gauge symmetries are realized in a universal manner, independent of the form of interactions \footnote{Even though this approach can access generic theories of the form \eqref{nlingen}, it would be interesting to check if there also exist theories that deform the symmetries \eqref{dashift} and \eqref{symashift} of the free theory.}.

The approach adopted here naturally lends itself to generalizations to higher form field interactions, and it would be interesting to explore them. An educated guess is that for $(2k-1)$-form fields in $4k$ dimensions, the most general nonlinear democratic Lagrangian is
\beq
{\cal L}={\cal L}_{free}+g(\lambda_i)\,,\label{pform}
\eeq
where the first term is the free Lagrangian from \cite{SOK} and the second term is an arbitrary function of all independent Lorentz scalars $\lambda_i$ built out of the tensor $\s H^a-\epsilon^{ab}H^b$. The equations of motion will imply a deformed twisted self-duality relation that expresses, as in (\ref{FF}),
$\s F^a+\epsilon^{ab}F^b$ as a function of the opposite chirality combination $\s F^a-\epsilon^{ab}F^b$.
The $SO(2)$ symmetry condition will further constrain $g$ to depend on a specific set of combinations of $\lambda_i$.
Similarly, for the chiral $2k$-forms in $4k+2$ dimensions, a natural guess is of the same form \eqref{pform}, with the free part given as in \cite{K,SOK}, while now $\lambda_i$ are all independent Lorentz scalars built out of the tensor $\s H-H$. 
These structures will be explored in more detail in future works.

\vspace{2mm}
\noindent{\bf Acknowledgments:}
The authors are grateful to Euihun Joung and Arkady Tseytlin for helpful discussions.
OE has been supported by the CUniverse research promotion project (CUAASC) at Chulalongkorn University. KM is supported by the European Union's Horizon 2020 research and innovation programme under the Marie Sk\l odowska-Curie grant number 844265.

\end{document}